\documentclass[conference]{IEEEtran}
\IEEEoverridecommandlockouts
\usepackage{cite}
\usepackage{amsmath,amssymb,amsfonts}
\usepackage{algorithmic}
\usepackage{graphicx}
\usepackage{textcomp}
\usepackage{xcolor}

\usepackage[pagewise,switch]{lineno}
\usepackage{algorithm}
\usepackage{url}
\usepackage[caption=false,font=footnotesize]{subfig}
\usepackage{cuted}          
\usepackage{fvextra}        

\DefineVerbatimEnvironment{WideVerb}{Verbatim}{
  breaklines, breakanywhere,
  frame=single, framerule=0.8pt, framesep=6pt, fontsize=\small
}

\def\BibTeX{{\rm B\kern-.05em{\sc i\kern-.025em b}\kern-.08em
    T\kern-.1667em\lower.7ex\hbox{E}\kern-.125emX}}

\begin{document}
\pagestyle{plain}
\title{LLM-Powered Code Analysis and Optimization for Gaussian Splatting Kernels\\
}

\author{\IEEEauthorblockN{Yi Hu}
\IEEEauthorblockA{
\textit{North Carolina State University}\\
Raleigh, USA \\
yhu34@ncsu.edu}
\and
\IEEEauthorblockN{Huiyang Zhou}
\IEEEauthorblockA{ 
\textit{North Carolina State University}\\
Raleigh, USA \\
hzhou@ncsu.edu}
}


\maketitle


\begin{abstract}
3D Gaussian splatting (3DGS) is a transformative technique with profound implications on novel view synthesis and real-time rendering. Given its importance, there have been many attempts to improve its performance. However, with the increasing complexity of GPU architectures and the vast search space of performance-tuning parameters, it is a challenging task. Although manual optimizations have achieved remarkable speedups, they require domain expertise and the optimization process can be highly time consuming and error prone. In this paper, we propose to exploit large language models (LLMs) to analyze and optimize Gaussian splatting kernels. To our knowledge, this is the first work to use LLMs to optimize highly specialized real-world GPU kernels. We reveal the intricacies of using LLMs for code optimization and analyze the code optimization techniques from the LLMs. We also propose ways to collaborate with LLMs to further leverage their capabilities. 


For the original 3DGS code on the MipNeRF360 datasets, LLMs achieve significant speedups, 19\% with Deepseek and 24\% with GPT-5, demonstrating the different capabilities of different LLMs. By feeding additional information from performance profilers, the performance improvement from LLM-optimized code is enhanced to up to 42\% and 38\% on average. In comparison, our best-effort manually optimized version can achieve a performance improvement up to 48\% and 39\% on average, showing that there are still optimizations beyond the capabilities of current LLMs. On the other hand, even upon a newly proposed 3DGS framework with algorithmic optimizations, Seele, LLMs can still further enhance its performance by 6\%, showing that there are optimization opportunities missed by domain experts. This highlights the potential of collaboration between domain experts and LLMs. Additionally, during the optimization process, we found that LLMs may sacrifice functional equivalence for speedups by introducing unsafe optimizations. We propose to address this critical issue with LLM cross checking, i.e., using one LLM to check the code optimized by another LLM for semantical equivalence with the unoptimized code. Among the LLMs used in our study, GPT-5 performs best in this task.   



\end{abstract}



\section{Introduction}


3D reconstruction technologies have promoted the development of computer graphics as well as computer vision, including navigation, virtual reality and extended reality. Neural Radiance Field approach, which leverage multilayer perceptron (MLP), were the landmark in the development of 3D rendering and regarded as the state-of-the-art approach to generate novel view frames of the scene which have not been captured by the camera \cite{muller2022instant}. Despite the acceleration techniques such as hash encoding and baking, it is hard for this volumetric rendering method to meet the requirement of real-time rendering due to the high cost of sampling and the expensive computation from MLP. 

Recently, a novel radiance field rendering technique called 3DGS has shown its capability to render real-time frame rates and has been regarded as the new state-of-the-art rendering approach. Instead of implicit embedding of the color and the density of a vertex and expensive multilayer perceptron, gaussian splatting conducts rasterization based on gaussian splats which store the color explicitly to simplify the computation in rendering. Additionally, there is no need for sampling in the rendering process because all the Gaussians have been generated during training. Gaussian splatting based rendering quickly attracts the community's attention and multiple optimization strategies have been explored. Some focus on software-level algorithmic optimization including computational workload reduction and better workload balanced parallelism \cite{gui2024balanced},\cite{hanson2025speedy},\cite{hollein20243dgs},\cite{huang2025seele}. Others adopt hardware-level acceleration which propose special purpose accelerators for Gaussian splatting rendering \cite{li2025uni},\cite{li2025gaurast}. 

Besides algorithmic optimizations, code optimization or better ways to implement the GPU kernels remain critical for performance. However, manually optimizing the 3DGS rendering pipeline requires a deep understanding of both the kernel workload and the system architecture. It is also time consuming and error prone due to the trial and error nature. On the other hand, researchers have explored the possibility of using Artificial Intelligence (AI) to optimize software code automatically \cite{hong2025autocomp},\cite{lange2025ai},\cite{novikov2025alphaevolve},\cite{zheng2020ansor}. In particular, the latest large language models (LLMs) \cite{achiam2023gpt},\cite{guo2025deepseek},\cite{hoffmann2022training},\cite{team2024gemini} have shown their strong capability to optimize GPU kernels such as general matrix multiplication (GEMM).

In this paper, we conduct analysis and optimization of the Gaussian splatting pipeline with the help of LLMs, aiming to simplify or possibly automate the process of analyzing and optimizing 3DGS kernels. In our study, we explore four LLMs, GPT-5 \cite{openai_gpt5_system_card_2025}, Deepseek r1\cite{guo2025deepseek}, Gemini\cite{team2024gemini}, Claude\cite{anthropic2024claude3}, and an iterative evolutionary search framework, Openevolve\cite{novikov2025alphaevolve}, on different 3DGS frameworks, including the original 3DGS \cite{hollein20243dgs}, Seele \cite{huang2025seele}, and TC-GS \cite{liao2025tc}. The latter two have both algorithmic- and code-level optimizations to accelerate 3DGS.  

Our results are summarized as follows. For the original 3DGS code on the MipNeRF360 datasets \cite{barron2022mip}, LLMs achieve significant speedups, 19\%(Deepseek) and 24\%(GPT-5), which showcases the different capabilities of different LLMs. By feeding additional information from performance profilers, the performance improvement increases to 42\% and 38\% on average with Deepseek. In comparison, our best-effort manually optimized version can achieve a performance improvement up to 48\% and 39\% on average, showing that there are still optimizations beyond the capabilities of current LLMs. On the other hand, upon the newly proposed 3DGS framework with both code and algorithmic optimizations, Seele, LLMs can still further improve its performance by 6\%, showing that there are optimization opportunities missed by domain experts. This highlights the potential of collaboration between domain experts and LLMs. For TC-GS, LLMs did not further improve the performance due to its use of the tensor core instructions.

During our code optimization process with LLMs, we make the following key observations and propose techniques to improve the efficacy of LLMs in code optimization.

\begin{enumerate}
\item{\textbf{Functional Equivalence}} We found that LLMs may sacrifice functional correctness or equivalence to the original code for higher speedups by introducing unsafe optimizations. We overcome this challenge by cross-referencing between LLMs, which means using one LLM to check the functional equivalence of the optimized code from another LLM with the unoptimized code. Among the LLMs in our study, GPT-5 performs the best in this task. However, although the checker LLM can detect functional un-equivalence, it may not be able to fix the problem, and we resort to manual intervention to correct the code when we optimize the Seele and TC-GS code.
\item{\textbf{Optimization Space Pruning}} Given the kernel source code, LLMs may suggest many possible ways to improve performance. Unfortunately, some of them may not be effective, which unnecessarily enlarges the search space for optimizations. To address this issue, we propose to feed the profiling data from performance profilers to LLMs so as to reduce their search space in optimizing the code.
\item{\textbf{Specialty vs. Generality}} We found that LLMs may over-optimize the code for a specific input. This may be desirable if the code is supposed to be used only on such inputs. If more stable performance of 3DGS is the goal, feedback from additional tests on different scenes would be helpful.
\end{enumerate}

In summary, we make the following contributions in this paper:

\begin{enumerate}
    \item We exploit LLM-based code optimization for 3DGS kernels.
    \item We analyze the optimization approaches proposed by LLMs.
    \item We identify the drawbacks in existing LLM-based code optimization and propose solutions to these drawbacks, which can be reused to optimize GPU kernels beyond 3DGS. 
\end{enumerate}

\section{Background}

\subsection{Overview of Radiance Field Rendering}

Radiance field rendering is a technique for generating realistic 2D images from 3D scenes by modeling the way light radiates through space. The representation of meshes and voxels are calculated from a series of views of the 3D scene by back propagation in the state-of-the-art methods. Voxel-based methods like Instant-NGP \cite{muller2022instant} use a small MLP to represent density and appearance and accelerate the ray marching by a hash grid and an occupancy grid. This kind of rendering achieves fast rendering and small memory consumption. However, MLP is still expensive in rendering. A new GPU friendly method, 3DGS \cite{kerbl20233d}, achieves $\sim 10$x frame rate compared to Instant-NGP and is likely to be the new standard for real-time rendering. 

\textbf{Accelerating 3DGS}
Given the importance of 3DGS, a series of work has been conducted to accelerate it. Uni-Render proposes a unified accelerator for mesh-based rasterization, NeRF, and 3DGS\cite{li2025uni}. GauRast extends the existing GPU triangle rasterizers for $\alpha-$ blending of Gaussian splats\cite{li2025gaurast}. Lee provides the characterization of 3DGS with Nsight systems and Nsight Compute, comparing it to Instant-NGP on open source datasets \cite{lee2024characterization}. Balanced 3DGS deals with the workload imbalance of thread blocks during training by dynamic workload allocation and Gaussian-wise parallelism\cite{gui2024balanced}. 4D-Rotor Gaussian Splatting represents dynamic scenes with anisotropic 4D XYZT Gaussians, demonstrating powerful capabilities for modeling complicated dynamics and fine details\cite{duan20244d}. 3DGS-LM replaces the ADAM optimizer with a tailored Levenberg-Marquardt and leverages a caching data structure for intermediate gradients to enable efficient calculation of Jacobian-vector products in custom CUDA kernels\cite{hollein20243dgs}. Speedy-Splat localizes the Gaussian in the image place and reduces the number of Gaussians by soft and hard pruning with a marginally decreased PSNR\cite{hanson2025speedy}. SeeLe leverages hybrid preprocessing and contribution-aware rasterization, which reduces the computation for low-frequency Gaussians, as well as double buffering to decrease the cost of synchronization \cite{huang2025seele}. TC-GS transforms the calculation of opacity into matrix multiplication and uses tensor cores to accelerate this process \cite{liao2025tc}.    

\subsection{Structure of Gaussian Splatting Rendering Pipeline}

3DGS \cite{kerbl20233d} is the state-of-the-art radiance field representation which leverages rasterization to perform high-quality rendering in real time. A general Gaussian Splatting rendering pipeline as shown in Algorithm~\ref{al:gs} can be divided into the following two stages. 

\textbf{(1) Preprocessing:}
All the trained Gaussians are loaded and only visible Gaussians in the current camera view are selected. Then 3D Gaussians are projected onto a 2D plane according to the pose of the camera, during which 2D splats with color and depth are generated. Next, these splats are replicated for each pixel tiles with a default size of 16 $\times$ 16. Then on each tile, overlapped splats are sorted based on the depth to maintain correct occlusion relationship for upcoming $\alpha$-blending, which is the most expensive part of the rendering. 

\textbf{(2) $\alpha$-blending:}
At first, bounds and ranges of splats for tile and pixel are prepared. Then, attributes of overlapped splats on a tile which is assigned to a thread block are loaded into the shared memory cooperatively. Next comes the pixel-wise blending. Density, transmittance, and color of the splats are calculated for each pixel in the order of current depth to the camera as shown in Equation~\eqref{eq:alpha_blending}. Then the final color of a pixel is derived from these attributes of bound splats. And early stop approach is adopted when the transparency falls below the threshold to improve the performance. 

\begin{equation}
    C = \sum^N_{i=1} T_i\alpha_ic_i
    \label{eq:alpha_blending}
\end{equation}

with 

$\alpha_i = (1-exp(-\sigma_i\delta_i))$, and $T_i=\prod\limits^{i-1}_{j=1}(1-\alpha_i)$,

where $\delta$ means the interval and $\sigma$ means the density. 

\begin{algorithm}
\caption{Gaussian Splatting Rendering Kernel}
\begin{algorithmic}[1]
\nonumber{ \STATE $\alpha$ : Gaussian's 2D splat opacity}
\nonumber{ \STATE $T$ : Gaussian's accumulated transmittance}
\nonumber{ \STATE $C$ : Color of the pixel rendered by the thread}
\STATE \textbf{Identify tile:} Compute bounds for tile and pixel. 
\STATE \textbf{Load work range:} Retrieve start/end indices; calculate number of batches.
\STATE \textbf{Allocate shared memory:} Reserve arrays for IDs, positions, and attributes.
\FOR{each batch}
    \STATE Fetch batch Gaussian data into shared memory synchronously.
    \FOR{each Gaussian in batch}
        \IF{thread not done}
            \STATE Compute $\alpha$, update $T$, $C$, other accumulators; check done flag.
        \ENDIF
    \ENDFOR
\ENDFOR
\IF{pixel is inside}
    \STATE Write $T$, $C$, contributor count, and optional outputs to global memory.
\ENDIF
\end{algorithmic}
\label{al:gs}
\end{algorithm}


\subsection{LLMs-based GPU Kernel Optimization}

LLMs have been used to optimize CUDA kernels. 
TVM auto-scheduler, as known as Ansor \cite{zheng2020ansor}, provides fine-tuned parameters, such as tile size and unroll factors. AI CUDA engineer \cite{lange2025ai}, proposes optimized CUDA kernels generated by special trained LLMs. Besides, GPT-5, Deepseek r1 and Gemini, also have the potential to generate high-quality CUDA kernels. LLMs support the AI agents to understand source code and search for optimization approaches. By learning from profiling data or simulation feedback, LLMs can often produce kernels that outperform hand-tuned baselines, significantly reducing development time while achieving near-peak hardware efficiency.

Considering an optimization $S$ of an input program $p_0 \in P$ for an objective function $f : P \to \mathbb{R}$, $S$ is a sequence of modification $\{m_i\}$, where $m_i \in M$ and each modification could change the performance while keeping the functional equivalence. The objective function $f$ is the evaluation of the program for a given set of merit (e.g.,
latency, accuracy, utilization). $P$ represents the set of functional equivalent variants of the target program. Finding an optimization $S$ can be formulated as 
\begin{equation}
    S = \arg\max\limits_{\lvert S \rvert \leq T} f((m_k \circ \cdots \circ m_1)(p_0))
\end{equation}
for a given maximum length of the sequence $T$. Each modification is generated by LLMs. A modification can be pointed out in the prompt based on expertise knowledge or produced by LLMs automatically. 

To maximize LLM's capacity and reduce the influence of randomness, existing work proposes architectures which can interact with LLMs by updating prompts based on different techniques and user feed back\cite{hong2025autocomp},\cite{lange2025ai},\cite{novikov2025alphaevolve},\cite{tang2025compiler}. These approaches exploit various search algorithms to find an optimal version of the target program in the modification space. Despite promising results, important gaps remain before these approaches are ready for real-world complex GPU kernels.  

\subsection{CUDA Kernel Performance}

This subsection presents a collection of the concepts and metrics from Nsight Compute (NCU) \cite{nvidia_nsight_compute_user_guide_2025} associated with GPU architecture and CUDA, the parallel computing platform and programming model developed by NVIDIA. These terms are fundamental to understanding the data that we feed to the LLMs. 

\textbf{Streaming Multiprocessor (SM):} An SM, containing multiple CUDA cores, is a major hardware component of the GPU architecture. Thread blocks are scheduled to SMs non-preemptively. The performance of a CUDA kernel is highly dependent on the efficiency of SMs.

\textbf{Stalls:} A stall occurs when  warps, the basic execution unit on a GPU, are unable to proceed with execution due to unmet conditions such as waiting for data from memory, contention for functional units, or unresolved dependencies. 

\textbf{Roofline model:} The roofline model, relating computational throughput to memory throughput as known as arithmetic intensity, is a common visual and analytical tool used to assess and optimize the performance of computational kernels on modern processors. Performance is plotted against arithmetic intensity on a log-log scale, forming a "roofline" shape. The sloped region of the graph represents the memory-bound regime, where performance scales with data bandwidth. The flat region indicates the compute-bound regime, where performance is capped by the processor’s maximum compute throughput.

\section{Capability of LLMs}

Existing LLMs can generate optimized code for 3DGS 
rendering kernels. In Table \ref{Improvement from LLMs}, we compare the runtime of different 3DGS frameworks, original 3DGS, Seele, and TC-GS, with the optimized versions generated by different LLMs, GPT-5, Deepseek r1, Gemini, and Claude. 

From Table \ref{Improvement from LLMs}, we can see that most LLMs in our study can optimize the original 3DGS rendering kernels by up to 24\% in one query without any additional information. However, for kernels that have been optimized manually, like those in Seele and TC-GS, the source code has higher complexity and there is a higher chance for LLMs to introduce errors in the generated code. For these kernels, LLMs failed to generate functionally correct optimized code and we have to fix the bugs manually in such buggy optimized code. An asterisk is added to the runtime in Table \ref{Improvement from LLMs} to denote such cases. 

\begin{table}[]
\centering
\caption{Execution time (ms) of different Gaussian splatting rendering kernels, orginal 3DGS, Seele, and TC-GS, optimized by LLMs for the scene, room, from MipNeRF 360\cite{barron2022mip}. 
}
\begin{tabular}{l|lll}
\hline
Version      & 3DGS          & Seele         & TC-GS     \\ \hline
Origin       & 4.71          & 1.26          & 0.43      \\ \hline
GPT-5        & \textbf{3.76} & \textbf{1.19*} & 0.44*      \\ \hline
Deepseek     & 4.01          & Failed        & Failed    \\ \hline
Gemini       & 4.34          & Failed        & Failed    \\ \hline
Claude       & Failed        & Failed        & Failed    \\ \hline
\end{tabular}
\label{Improvement from LLMs}
\end{table}

\begin{figure}[htbp]
    \centering
    \includegraphics[width=0.48\textwidth]{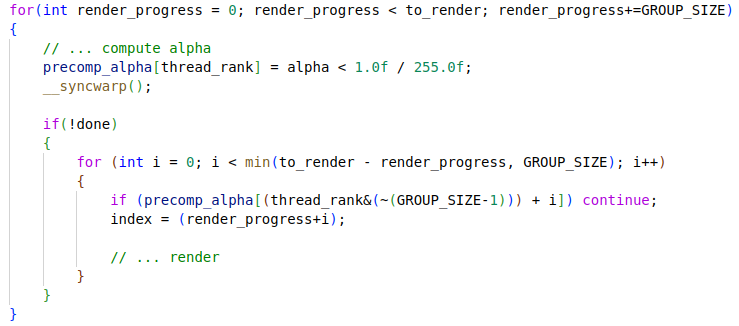} 
    \caption{Original version of blending in Seele\cite{huang2025seele}}
    \label{Comparision of original seele}
    \includegraphics[width=0.48\textwidth]{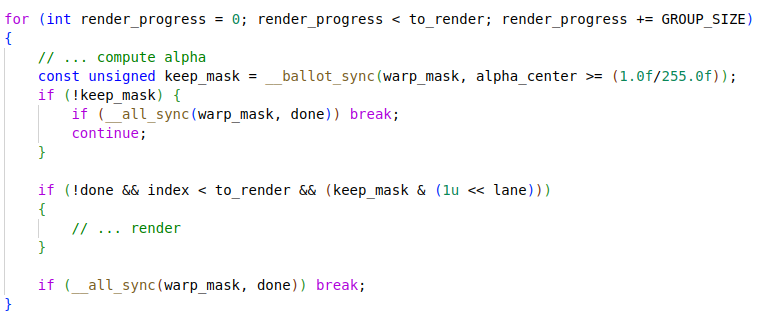} 
    \caption{GPT-optimized version of blending in Seele}
    \label{Comparision of GPT seele}
    \includegraphics[width=0.48\textwidth]{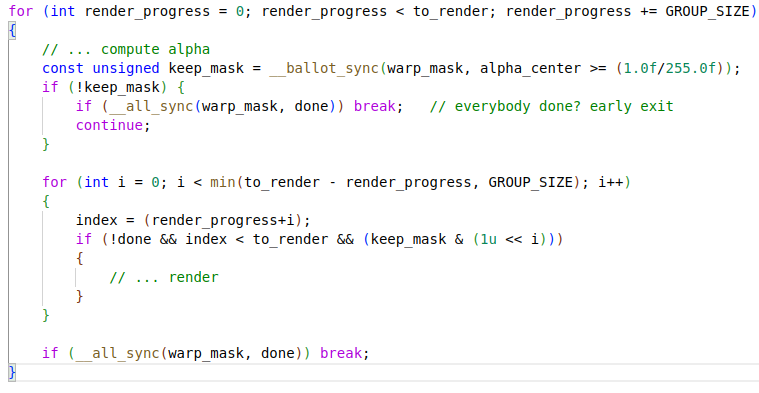} 
    \caption{Manually corrected GPT-optimized version of blending in Seele}
    \label{Comparision of manually corrected}
\end{figure}

Next, we use the optimization of Seele as a case study. The original version uses shared memory to generate and share masks in a thread block, as shown in Figure \ref{Comparision of original seele}. Given the size of a group can be covered by a warp, GPT-5 proposes to use \textit{\_\_ballot\_sync}, a hardware primitive which associates register transfer and synchronization within a warp, instead of shared memory. This is a parameter-sensitive optimization as it only works for certain group size (4*8). However, from the code generated by GPT-5 shown in Figure \ref{Comparision of GPT seele}, we can see the generated code does not keep the functional equivalence compared to the original version. GPT-5 treats the inner loop as redundant computation and removes it, which causes a severe loss of the quality of the rendered images. We fix the bug manually as shown in Figure \ref{Comparision of manually corrected}. With our manual fix, the optimized code from GPT-5 gains a 6\% speedup compared to the original Seele code.  

From this case study, we can see that LLMs can discover some optimization opportunities that may be missed by domain experts. However, they may fail to guarantee the functional equivalence of the generated code, which is an obstacle to overcome. 

\subsection{Dissecting Optimization Techniques from LLMs}

We analyze the optimized code from LLMs and categorize the optimization techniques as follows: 

\begin{itemize}
    \item \textbf{Reduce computational overhead.} LLMs replace mathematical functions with the optimized version. For example, the \textit{exp} function is replaced with \textit{$\_\_$expf}. 
    \item \textbf{Reduce redundant variables.} LLMs can analyze the logic of the source code and reduce some redundant variables by simple modifications. 
    \item \textbf{Optimize loop organization.} LLMs can help redesign kernel loops to reduce warp divergence and streamline control flow.
    \item \textbf{Memory coalescing. } LLMs can identify coalescing opportunities in the code, which can improve the memory access efficiency.  
    \item \textbf{Shared memory.} LLMs can find redundant global memory accesses and put frequently accessed data into shared memory to improve data reuse.
    \item  \textbf{Warp-level primitives.} LLMs can use warp-level primitives for fine-grained parallelism and better resource utilization.  
\end{itemize}

Next, we provide more details on the optimizations that LLMs apply on the original 3DGS kernel.

As the original 3DGS kernel has a high arithmetic intensity, the optimization of using \textit{$\_\_$expf} to reduce the computation of $\alpha$ of a Gaussian results in a 10.35\% improvement. 

The 3DGS code has a counter, \textit{contributor}, which can be removed by calculating it directly with existing iteration counters. This way, the kernel can get a 0.91\% improvement.  

Simplifying the loop condition computation can enable the compiler to unroll the loop. The original code has a flag which might be changed in the loop body. This prevents potential unrolling by the compiler. After simplifying the loop condition, the compiler unrolls two iterations by default, which results in a 5.66\% improvement. 

A coalesced load of RGB channels can reduce the times of global loads, which provides a 3\% improvement. 

Optimizing the memory layout by putting shared variables in the shared memory can improve the memory access efficiency, which can result in a 11.3\% improvement. However, LLMs do not do it automatically. Although LLMs have the capability to find the global memory access pattern, this kind of optimization still needs to be specified in the prompt. 

Warp-level primitives allow direct data exchange between registers of threads in the same warp, which avoids memory traffic and synchronization barriers, reducing latency.

\begin{figure*}
    \centering
    \includegraphics[width=1.0\linewidth]{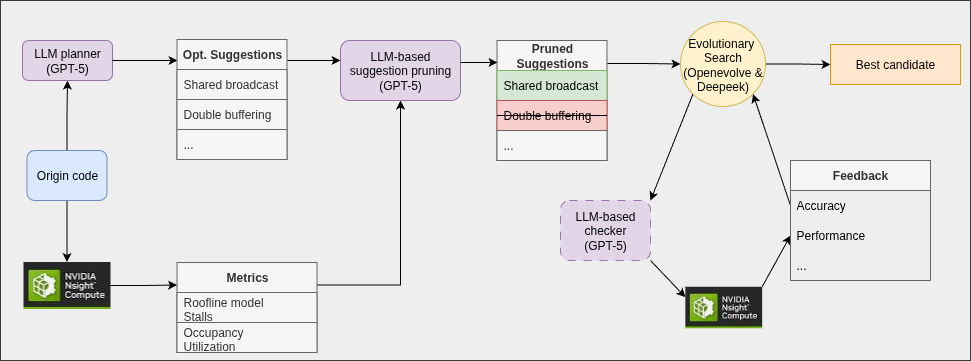}
    \caption{LLM-powered code optimization workflow. One LLM serves as the planner and prunes the search space of combined optimizations with profile data which characterize the workload. And we adopt an Evolutionary Search engine for code generation with LLM-powered correctness check on the generated code.}
    \label{architecture}
\end{figure*}

In summary, there are various optimizations that could affect the performance of the target workload, which introduces a high dimension search space of combined optimizations. And the complexity of the 3DGS kernels also challenges the LLM to apply optimizations correctly and properly.


\section{How to Better Collaborate with LLMs } \label{collabllm}

\subsection{Key Challenges}

Although existing commercial LLMs can optimize 3DGS rendering kernels to a certain extent with simple prompts, there exist challenges to be addressed to better harness their capabilities. 

\begin{figure}[htbp]
    \centering
    \includegraphics[width=0.48\textwidth]{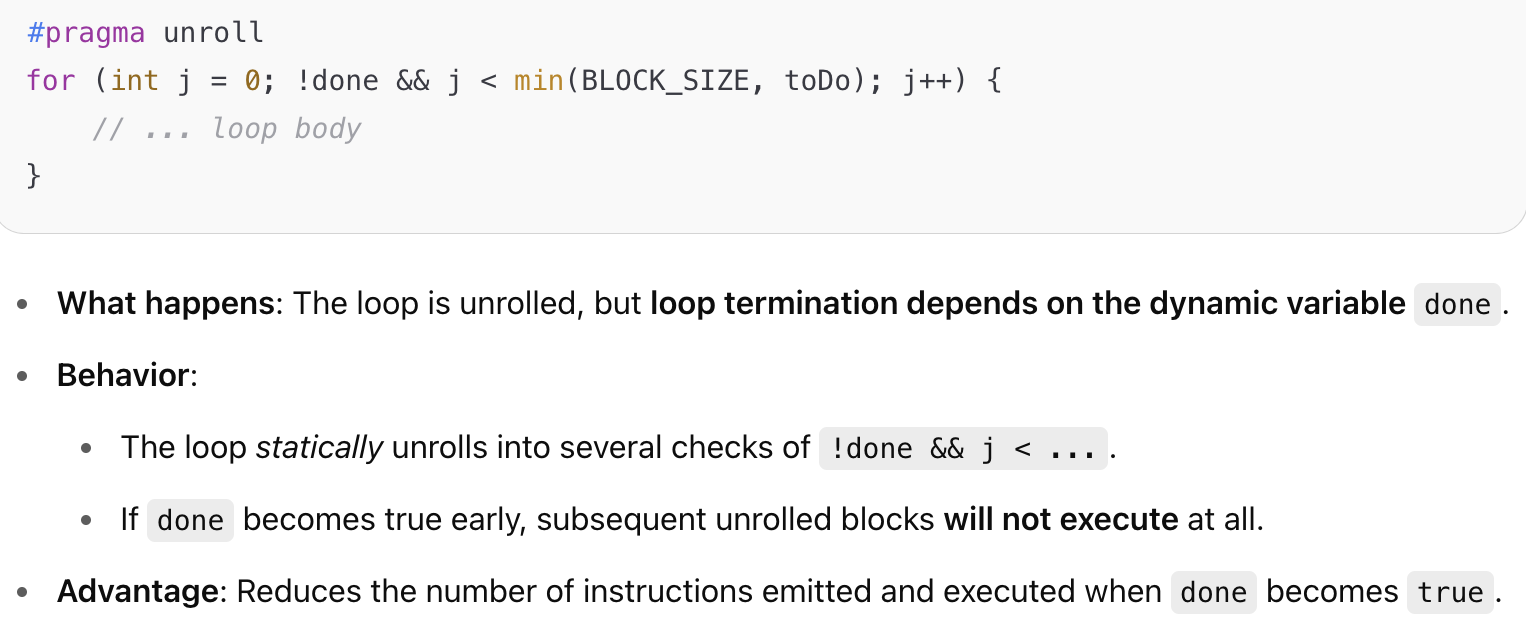} 
    \caption{Source code of an inner loop in the 3DGS kernels. If \textit{done} flag is changed to \textit{true} at early stage in the loop body, this implementation can reduce unnecessary iterations.}
    \label{Comparision of two loop_1}
    \includegraphics[width=0.48\textwidth]{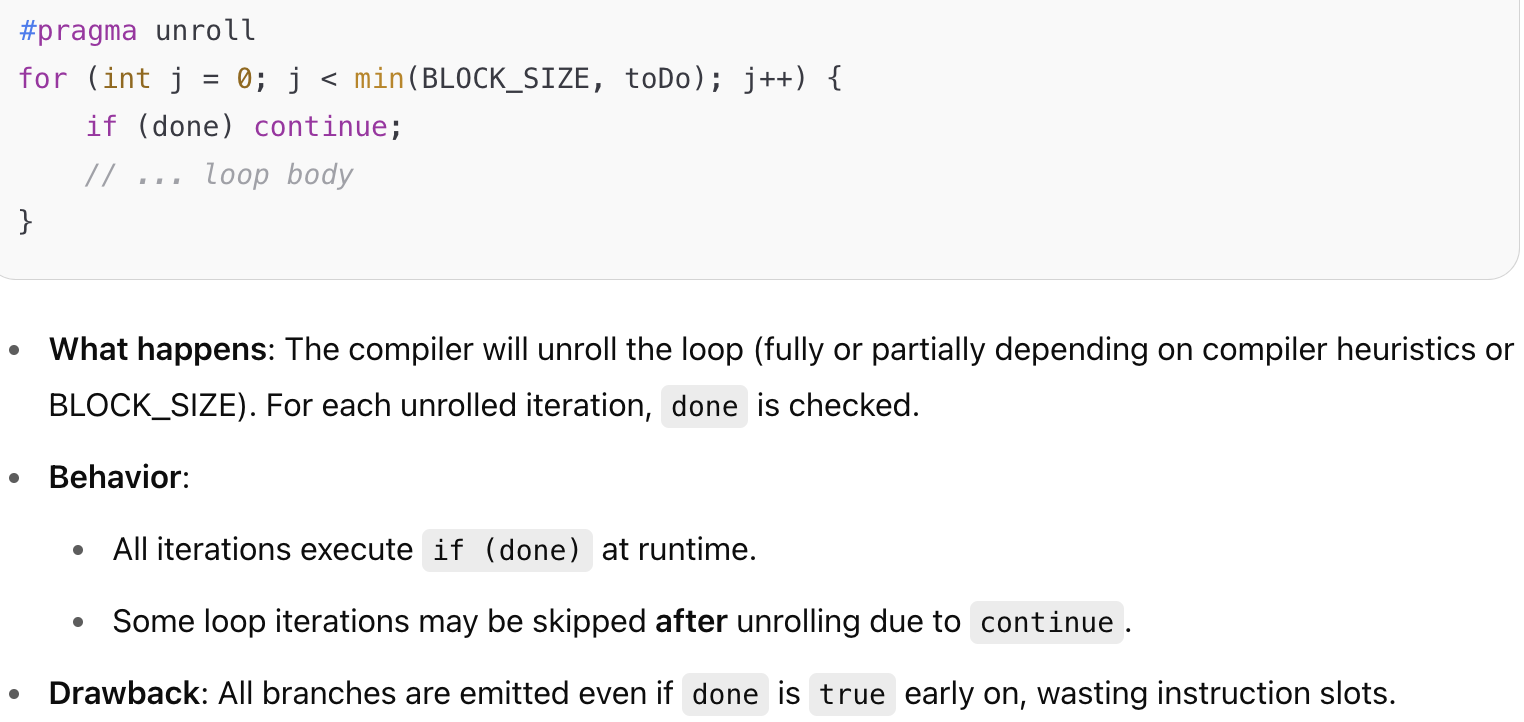} 
    \caption{LLM optimized version of the inner loop in Fig. 5. This implementation enables loop unrolling and achieves performance improvement when the number of iterations is high.}
    \label{Comparision of two loop_2}
\end{figure}

\begin{itemize}
    \item \textbf{Effectiveness.} LLMs provide a series of optimizations with corresponding conditions. Without the information on certain aspects like the input, LLMs can not tell whether these conditions are satisfied. Therefore, the resulting code may not improve the performance and may even cause performance degradation. One such example is shown in Figure \ref{Comparision of two loop_1} and Figure \ref{Comparision of two loop_2}, the optimized version has better performance than the un-optimized one only when the number of iterations is high.

   \item\textbf{Complexity} Considering the complexity of the search space of combined optimizations, it takes a large number of trials to search the space, which increases the difficulty to find the optimal implementation.  
    \item \textbf{Functional Equivalence.} Depending on the training material of these LLMs, they may not learn some characters or expressions in the provided code. Hence, they might discard these unknown contexts, which changes the functionality of the kernel and results in lower accuracy or even errors. Furthermore, splatting kernels are fairly complex, 
    the use of primitives, conditions, loops, and memory layouts make it difficult for LLMs to generate functional equivalent code. For example, we found that LLMs might choose to simplify the calculation by removing calculations which they think redundant while in fact not as demonstrated in the previous case study. 
  
\end{itemize}


\begin{figure*}[!t]
  \centering
  \subfloat[Advice 1-5]{\includegraphics[width=.49\textwidth]{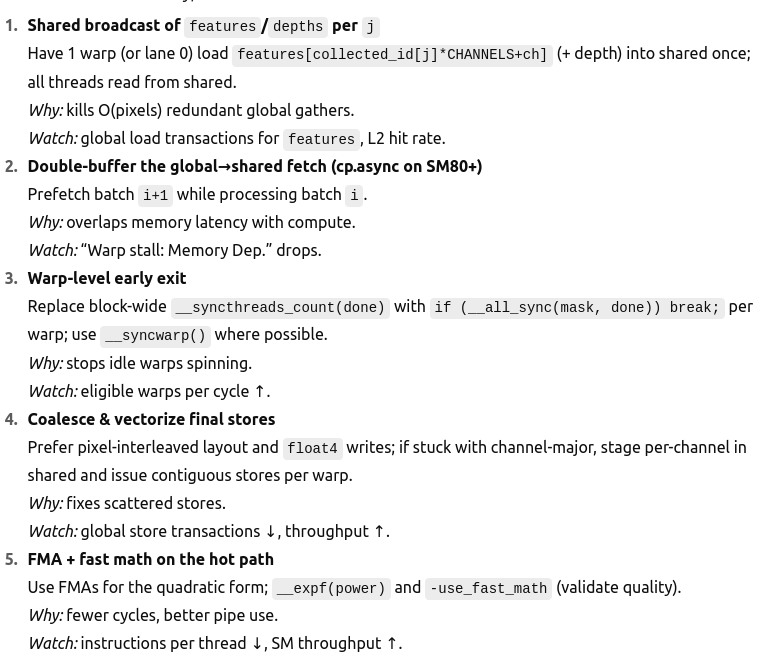}\label{plan1}}
  \hfil
  \subfloat[Advice 6-10]{\includegraphics[width=.49\textwidth]{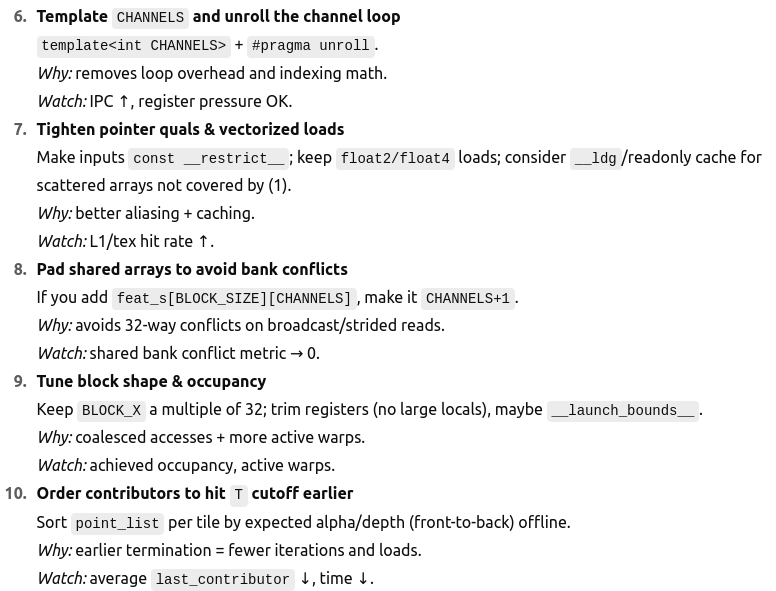}\label{plan2}}
  \caption{Advices from the planner, GPT-5, with the source code as input.}
  \label{plan}
\end{figure*}

To deal with these challenges, we propose the following solutions. The overall workflow of utilizing the LLMs for code optimization is shown in Figure \ref{architecture}.





\subsection{Solution 1: Adopting LLM as a Planner}

Previous studies have shown the complexity of code optimization makes it difficult for LLMs to produce semantically and syntactically valid programs \cite{agarwal2024codemirage},\cite{wang2024large}. Besides, the search space of combinations is of high dimension, as there are various optimizations and for each optimization the implementation varies. Therefore, a series of work shows that decomposing tasks into multiple steps can enhance the capability of a LLM in solving a complex task\cite{gao2023pal},\cite{hong2025autocomp},\cite{hong2024llm}. Hence, we use one LLM to analyze the source code and provide potential optimizations to guide code generation.  These optimizations are in plain language so that human developers can audit and adjust. These suggested optimizations can also inspire human developers for future development.
As shown in Figure \ref{plan}, GPT-5 proposes a series of optimizations based on the source code. We include these suggestions in the prompt for subsequent LLMs to guide their code optimization process.

\subsection{Solution 2: Pruning Search Space with Profiling Data}

As the planner may provide a large number of possible optimizations, trying all of them and their combinations would be too costly. We propose to provide the planner LLM with workload characteristics like the runtime information of the unoptimized code such that it can make judicious judgement on which optimizations are more likely to be effective. 
The following two sets of profiling data are provided:

\begin{itemize}
    \item \textbf{System Information.} These include profile results from Nsight, Nsight Compute, and logs from simulators. 
    \item \textbf{Dataset/Workload Distribution.} These include how the workload is distributed among thread block and threads. 
\end{itemize}

\subsection{Solution 3: Search-based Code Generation}

Even after pruning, the implementation and code generation remain complex due to the combinations of optimizations and various parameters. Hence, existing work adopts search algorithm to find the optimal implementation, including Evolutionary Search\cite{novikov2025alphaevolve}, Monte Carlo tree search\cite{tang2025compiler} and Beam search\cite{hong2025autocomp}. In this paper, we focus on Evolutionary Search. We build the evaluator based on two criteria \textit{accuracy} and \textit{performance}:

\begin{itemize}
    \item \textbf{Accuracy.} After each iteration of code generation, the generated code is compiled and run with our functional test suite. We verify the output of the candidate kernel against the original kernel on one scene. 
    \item \textbf{Performance.} We measure the latency of the candidate via Nsight Compute. 
\end{itemize}

These two metrics form the combined score as the return value of the evaluator. We record the best (highest combined score) candidate over iterations.  

\subsection{Solution 4: LLM-Powered Correctness Check}


From our experiments, LLMs are not reliable in generating optimized code for a complex kernel as they might not provide a functional equivalent modification. Besides, they may occasionally generate garbled texts outside the marked block where the optimizations are supposed to happen.
These errors degrade the quality of generated code and manually correcting is not practical in this search-based code generation process. Existing work exploits LLMs for code repair \cite{bouzenia2024repairagent},\cite{wadhwa2024core}. 
But they failed in fixing the erroneous code resulting from optimizing Seele and TC-GS code. Therefore, rather than fixing the errors, we choose to ask an LLM to check the functional equivalence of the optimized code against the original code. In our experiments, GPT-5 successfully accomplishes the task although it cannot fix the difference. 


\section{Methodology}

In this section, we present the system information and the characterization of the Gaussian splatting kernels. This information is provided to the planner LLM for pruning the suggested optimizations as discussed in Section \ref{collabllm}.

\subsection{Characterization Setup \& Datasets}

\textit{Datasets:} We utilized static real-world scenes from Mip-NeRF360\cite{barron2022mip} and Dr Johnson\cite{drjohnson_webpage} for 3DGS. All models were trained for 7000 iterations. Images from Mip-NeRF360 were downscaled by 4 to reduce VRAM usage. Images from Dr Johnson were downscaled by 2.

\textit{Hardware \& Software:} Our evaluation platform is an NVIDIA RTX 4060 GPU 
with 8GB VRAM, hosted with an Intel i5-13400F processor. It has 24 SMs and each SM can hold 48 
warps, that is, 6 = 48*32/(16*16) thread blocks of size [16,16]. CUDA version is 12.2 and PyTorch version is 1.12. We use Nsight Compute to characterize the workload.  

\textit{Gaussian kernels:} We use the 3 different versions of Gaussian splatting kernels, the original 3DGS \cite{hollein20243dgs}, Seele \cite{huang2025seele}, and TC-GS\cite{liao2025tc}.

\textit{LLMs:} We explore GPT-5 \cite{achiam2023gpt}, Deepseek r1 \cite{guo2025deepseek}, Claude \cite{anthropic2024claude3} and Gemini \cite{team2024gemini} to optimize the Gaussian splatting kernels. The Evolutionary Search is from Openevolve\cite{novikov2025alphaevolve}. We use Deepseek-reasoner API as the backend LLM to generate the optimized code and GPT-5 as the planner. The reason for such a choice is the cost of the APIs.

\subsection{System Information}

Metrics from Nsight Compute help to demystify GPU behavior into actionable insight, which shows the utilization of units and suggests the potential bottlenecks. Such information can help explain why the proposed optimization from LLMs would work, and select optimizations that have higher possibility to achieve performance gains. In this paper, we mainly focus on the following metrics: 
 
\begin{enumerate}
    \item \textbf{Roofline model:} We include the arithmetic intensity and performance of the turning point (or the knee of the Roofline curve) of the workload. Gaussian splatting kernel only uses float32. This information can help determine whether the workload is compute- or memory-bound. 
    \item \textbf{Stalls:} We include the composition of stalls which can expose the hidden latency in memory, control flow, and pipeline scheduling. This information helps determine which part should be optimized first.
    \item \textbf{Occupancy:} We include the occupancy of SMs which provides key insights into how a GPU kernel uses the underlying hardware. It reveals the constraints from resource allocation, exposes latency hiding potential, and offers information on tuning the register and shared memory usage. 
    \item \textbf{Utilization:} We include the unit with the highest utilization. Unit utilization offers a direct lens into the execution efficiency of GPU kernels. It helps identify whether the kernel is compute-bound, memory-bound, or limited by special functional units. This information is useful for optimizations such as loop restructuring, tiling, instruction reordering, or math function simplification.
\end{enumerate}

\begin{table}[h]
\centering
\caption{Attributes of system information collected by Nsight Compute of 3DGS.}
\begin{tabular}{p{3.2cm}p{1.4cm}p{1.4cm}}
\hline
Attributes                                           & Dr Johnson & MipNeRF360 \\ \hline
Arithmetic intensity of turning point (FLOP/byte)    & 42.63      & 42.63      \\ \hline
Performance of turning point (1e+12 FLOP/s)          & 11.24      & 11.24      \\ \hline
Arithmetic intensity of GS kernel (FLOP/byte)        & 253.68     & 235.35     \\ \hline
Performance of GS kernel (1e+12 FLOP/s)              & 2.34       & 2.22       \\ \hline
Warp cycles per issued instruction (cycle)           & 12.88      & 12.95      \\ \hline
Theoretical  occupancy (\%)                          & 100        & 100        \\ \hline
Achieved occupancy (\%)                              & 94.77      & 95.25      \\ \hline
Block limit warps (block)                            & 6          & 6          \\ \hline
Stall not selected                                   & 4.55       & 4.21       \\ \hline
Stall wait                                           & 2.59       & 2.59       \\ \hline
Stall barrier                                        & 0.91       & 1.45       \\ \hline
Stall short scoreboard                               & 1.19       & 1.18       \\ \hline
Selected                                             & 1.00       & 1.00       \\ \hline
Stall math pipe throttle                             & 0.95       & 0.85       \\ \hline
Unit with highest pipe utilization of active cycles  & ALU(57.1)  & ALU(57.1)  \\ \hline
\end{tabular}
\label{system info}
\end{table}

The system level information we collected on 3DGS is reported in Table \ref{system info}. 
The arithmetic intensity of Gaussian splatting kernels is higher than the turning points, which implies Gaussian splatting kernel is compute bound. In the current pixel-level parallelism, Gaussian splats that are applied to a thread block are loaded to the shared memory from the global memory cooperatively with the size of the thread block each time. Hence, compared to the computation of transmittance and color, memory accesses are faster, resulting in low L2 cache throughput and DRAM throughput. The occupancy of SMs are high. However, this depends on the platform. In our experimental setup, we use an NVIDIA 4060 GPU which has 24 SMs, each of which can hold 2048 threads concurrently. Considering an image with 778*519 pixels, there are 49*33 = 1617 thread blocks of size 16*16. The occupancy limit due to block size is 6, for which 144 (=6*24) thread blocks can be launched concurrently on this particular GPU. Then the GPU needs to run $\lceil1617/144\rceil=12$ waves, where a wave refers to a group of thread blocks that are executed concurrently. A large number of waves can provide flexibility for the scheduler to deal with inter-block imbalance. If the same task is launched on an A100 which has 108 SMs, there will be no more than 2.5 waves, in which case inter-block balance becomes important according to Balanced 3DGS\cite{gui2024balanced}. This can lead to different performance issues of 3DGS rendering on high-end server GPUs vs. edge devices. Therefore, it is essential to provide the system-level information to LLMs for platform specific optimizations. 
As the achieved occupancy is high, the stalls for 'not being selected' is the dominant reason, which is expected. 






\subsection{Workload distribution}

Workload distribution describes the workload assigned to thread blocks, warps and threads, which can help determine the existence of imbalance. In this paper, we collect the following metrics,

\begin{enumerate}
    \item \textbf{The distribution of splats for each tile:} We uploaded the distribution of splats for each tile. This information can reflect the workload assignment to each thread block in case of the inter-block workload imbalance.
    \item \textbf{The distribution of splats calculated for each thread in a tile:} We calculated the proportion of the splats calculated for each thread against the number of assigned splats. This information can reflect the influence of the early stop method in case of the intra-block workload imbalance.  
\end{enumerate}


\begin{table}[h]
\centering
\caption{Workload distribution among thread blocks, warps and threads.}
\label{workload}
\begin{tabular}{p{3.2cm}p{1.4cm}}
\hline
Average of Gaussians for each tile               & 1189                    \\ \hline
Variance of Gaussians for each tile              & 614608                  \\ \hline
Average of calculated Gaussians per thread(\%)   & 95                      \\ \hline
Variance of calculated Gaussians per thread      & 0.02                    \\ \hline
\end{tabular}
\end{table}

As shown in Table \ref{workload}, the variance of Gaussians for each tile is high, which implies high inter-block imbalance. A large number of waves is necessary to cover the imbalance of workload on each SM. Also, we can see a low variance of calculated Gaussians for each thread, meaning the intra-block imbalance is low. And most of the Gaussians are calculated. In such cases, the early stop approach will not gain much benefit. By removing early stop approach, the 3DGS kernel gets a little improvement, which is captured in one version of generated code, while for Seele, this trick does not work. 




\subsection{Planning and Pruning}

\begin{figure}
    \centering
    \includegraphics[width=0.9\linewidth]{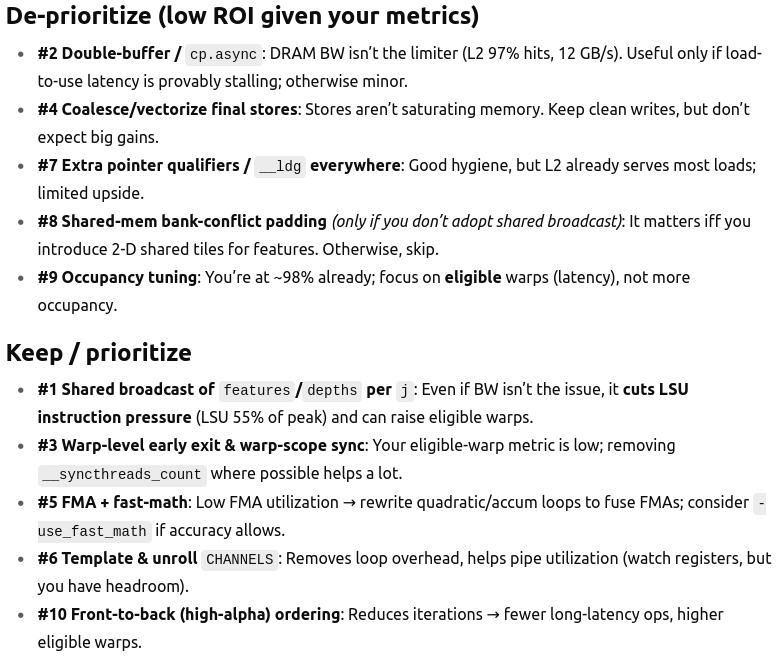}
    \caption{Selection from the suggested optimizations by the planner LLM, GPT-5, based on profile data. GPT-5 discards suggestions that are less likely to produce a speedup.}
    \label{plan_prune}
\end{figure}

LLMs can provide optimization suggestions based on source code, even though not all the suggestions are suitable on a target platform. As shown in Figure \ref{plan_prune}, by providing metrics profiled on the target platform, the planner LLM can decide and select the most promising suggestions, which reduce the dimension of the search space and the complexity of code generation.  

\subsection{Evolutionary Search}

For Evolutionary Search from Openevolve\cite{novikov2025alphaevolve}, we use Deepseek-reasoner as the backend LLM model which conducts code generation and quality-diversity evaluation. As for the performance of the generated kernel, we evaluate it using accuracy, which represents the distance between the output of the original kernel and that of the generated kernel, and the elapsed time measured by Nsight Compute. The suggestions from the planner are included in the prompt message to the LLM, and the examples are shown in the appendix.  


\subsection{Different version of 3DGS}
There have been proposals to improve upon the original 3DGS \cite{hollein20243dgs}. Seele \cite{huang2025seele} cuts redundant work via pruning and contribution-aware scheduling, which suits mobile and other resource-constrained GPUs. TC-GS \cite{liao2025tc} maps per-pixel alpha blending to GEMMs on NVIDIA Tensor Cores and uses FP16-safe transforms to keep fidelity stable. These manual modifications have introduced new features and transformed the search space of the combined optimizations. For instance, with accelerated opacity calculation on Tensor Cores, TC-GS has reduced the computational overhead a lot and it has become memory-bound. 


\section{Evaluation}






\subsection{Setup}

We pick one image \textit{room} from  MipNeRF360 as the test case for Evolutionary Search of the optimal version of 3DGS kernel. Considering one 3DGS kernel renders one image at a time and running that kernel in the simulator takes more than 10 minutes, we do not evaluate the generated code on full views. Evaluating generated codes on the whole benchmark grants better generality, although time-consuming. 

\textbf{Objective function. } The objective function for Evolutionary Search is a combined metric of efficiency measured by Nsight Compute and accuracy measured by the distance between ground truth and the output.

\textbf{Target workload. } The complexity of Seele and TC-GS leads to LLMs generating semantically and synthetically incorrect code, which require manual intervention. Therefore, we choose not to use search-based code generation for them. 
Hence, we use LLM-generated codes of Seele as the target for correctness check. And we evaluate the speedup and generality of LLM-generated codes based on 3DGS. 

\textbf{GPU architecture. } We conduct the Evolutionary Search on the same platform (i.e., RTX 4060) where we collect the system metrics. And we further evaluate the speedup of generated codes on three different GPUs (i.e., RTX 2060super, RTX A4000, and RTX A5000). 


\subsection{Runtime Comparison}

We use the same platform that we use for workload characterization. 

\subsubsection{Correctness check}

\begin{table*}[h]
\centering
\caption{Cross referencing results of functional equivalence of the generated code vs. unoptimized code. Modifications are generated by four LLMs based on Seele separately and all four versions fail to keep equivalence.}
\label{correctness}
\begin{tabular}{p{3cm}p{2.8cm}p{2.8cm}p{2.8cm}p{2.8cm}}
\hline
LLMs as checker    & GPT-5 version & Deepseek\_r1 version & Gemini version & Claude version \\ \hline
GPT-5             & Yes                & Yes                  & Yes            & Yes    \\ \hline
Deepseek\_r1           & No                 & Yes                  & Yes            & No     \\ \hline
Gemini                 & No                 & No                   & Yes            & No     \\ \hline
Claude                 & No                 & Yes                  & No             & No     \\ \hline
\end{tabular}
\end{table*}

For complex code like Seele, directly asking LLMs to optimize it might change the functional equivalence. Hence it is important to check the generated code. Different AI models show different levels of intelligence. Table \ref{correctness} illustrates the capability of different LLMs in identifying the inequality in the generated code, where a \textit{yes} represents that LLM can find the inequality in the generated code. GPT-5 can figure out the improper modifications in all these four versions while others cannot. Therefore, we recommend GPT-5 for correctness check.  

\subsubsection{Pruning}

\begin{figure}
    \centering
    \includegraphics[width=0.9\linewidth]{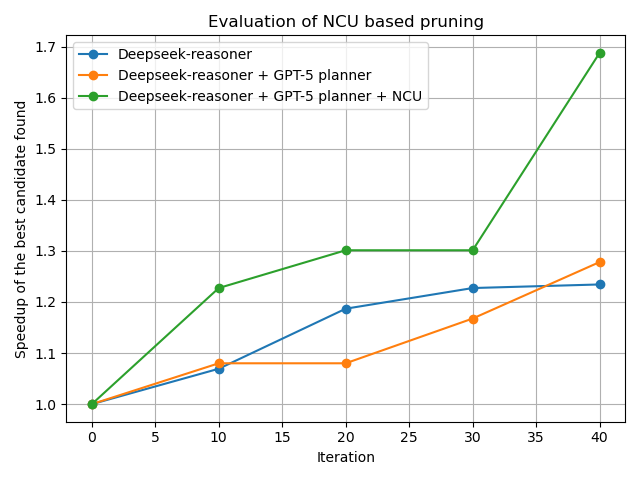}
    \caption{Speedup achieved through Evolutionary Search. The code generator and reviewer is Deepseek-reasoner, with GPT-5 as planner and NCU as profile tool. } 
    \label{prune}
\end{figure}

\textbf{Speedup relative to iterations.} Figure \ref{prune} shows the speedup over \textit{Evolutionary Search-Generated Code}, against the number of iterations on the x-axis. Speedup represents the ratio between the best code from Evolutionary Search and the original code. We record the speedup of the best candidate found every 10 iterations. For basic Evolutionary Search, the gains drop gradually, followed by convergence, and the search space of optimization is not fully exploited. When the suggestions from GPT-5 are added to the prompt, the LLM tends to exploit more combinations of optimizations and the gains grow slowly with some rapid rises. With a larger search space to exploit, the cost will increase, although there is possibility to find a better solution. In comparison, after pruning the search space with metrics from Nsight Compute, the gains grow much faster, which shows precise suggestions can help identify high rewarding regions in the search space. 

\begin{figure}
    \centering
    \includegraphics[width=0.9\linewidth]{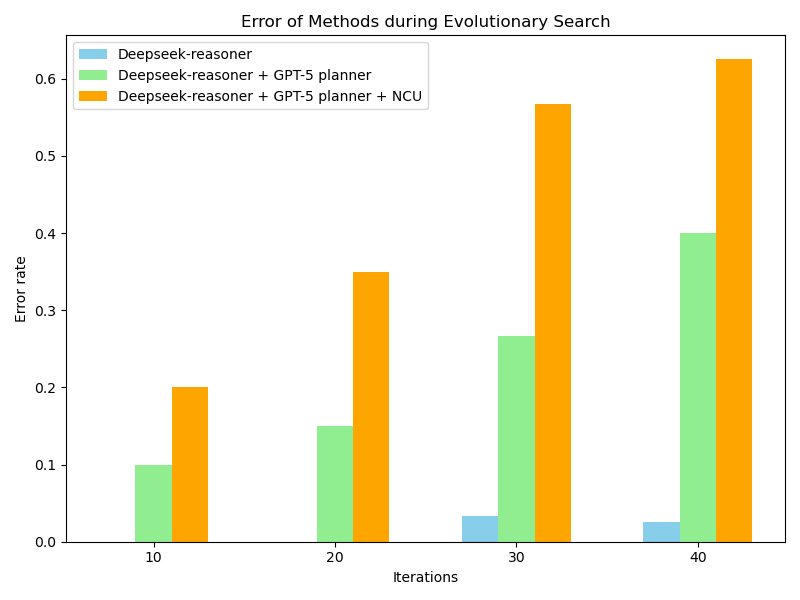}
    \caption{Error rate of search-based code generation over iterations.}
    \label{prune_error}
\end{figure}

\textbf{Error rate. } Figure \ref{prune_error} shows the error rates against the number of iterations on the x-axis. Here, an error means the generated code fails to compile or run to completion. Adding suggestions increases the complexity of the task and the LLM has a higher chance to produce errors in the generated code. Without correctness check, lots of iterations are wasted even though a better solution may be found eventually, which significantly increases the cost of API queries. In the Evolutionary Search, the LLM is called two times in each iteration for code generation and quality-diversity evaluation. If the error rate is higher than 1/3, then adding one query for correctness check would be beneficial.

\subsubsection{Generality}


\begin{figure*}
    \centering
    \includegraphics[width=0.8\linewidth]{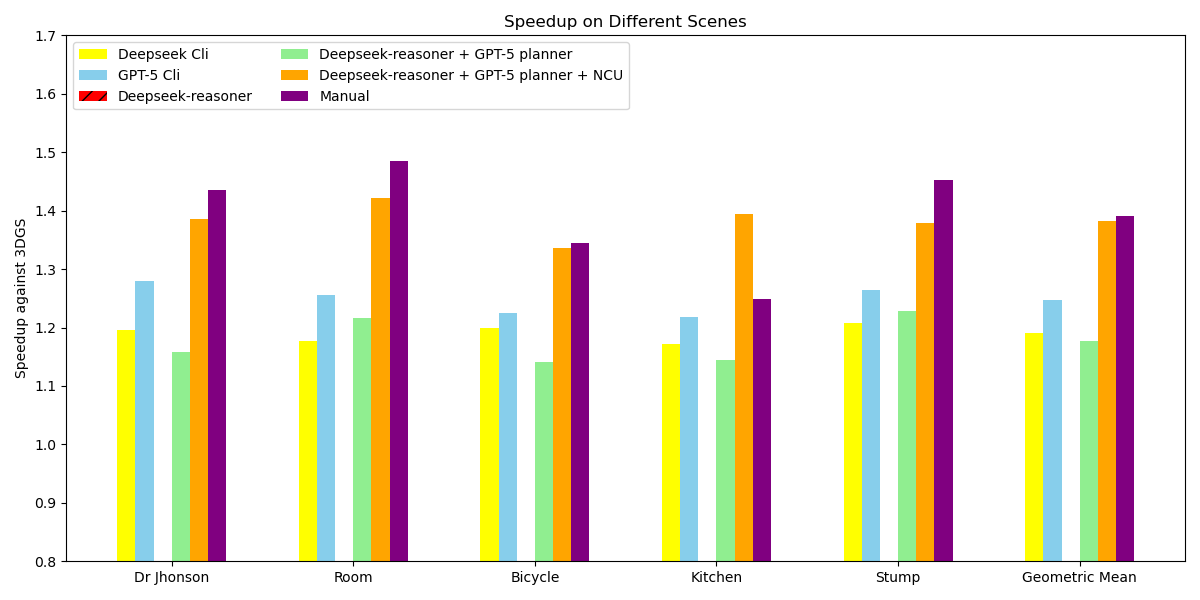}
    \caption{The speedups achieved from LLM generated optimized code over the original 3DGS kernel on different scenes. Deep-reasoner (without planner) failed to generate correct code for all the scenes. Deepseek/GPT-Cli means LLM-generated code without scene-specific information; Manual means our best-effort manually optimized code; the rest are search-based LLM optimized code with GPT-5 as planner and Deepseek as reasoner with \& without profile data from NCU.}
    \label{general_3dgs}
\end{figure*}

\begin{figure*}
    \centering
    \includegraphics[width=1\linewidth]{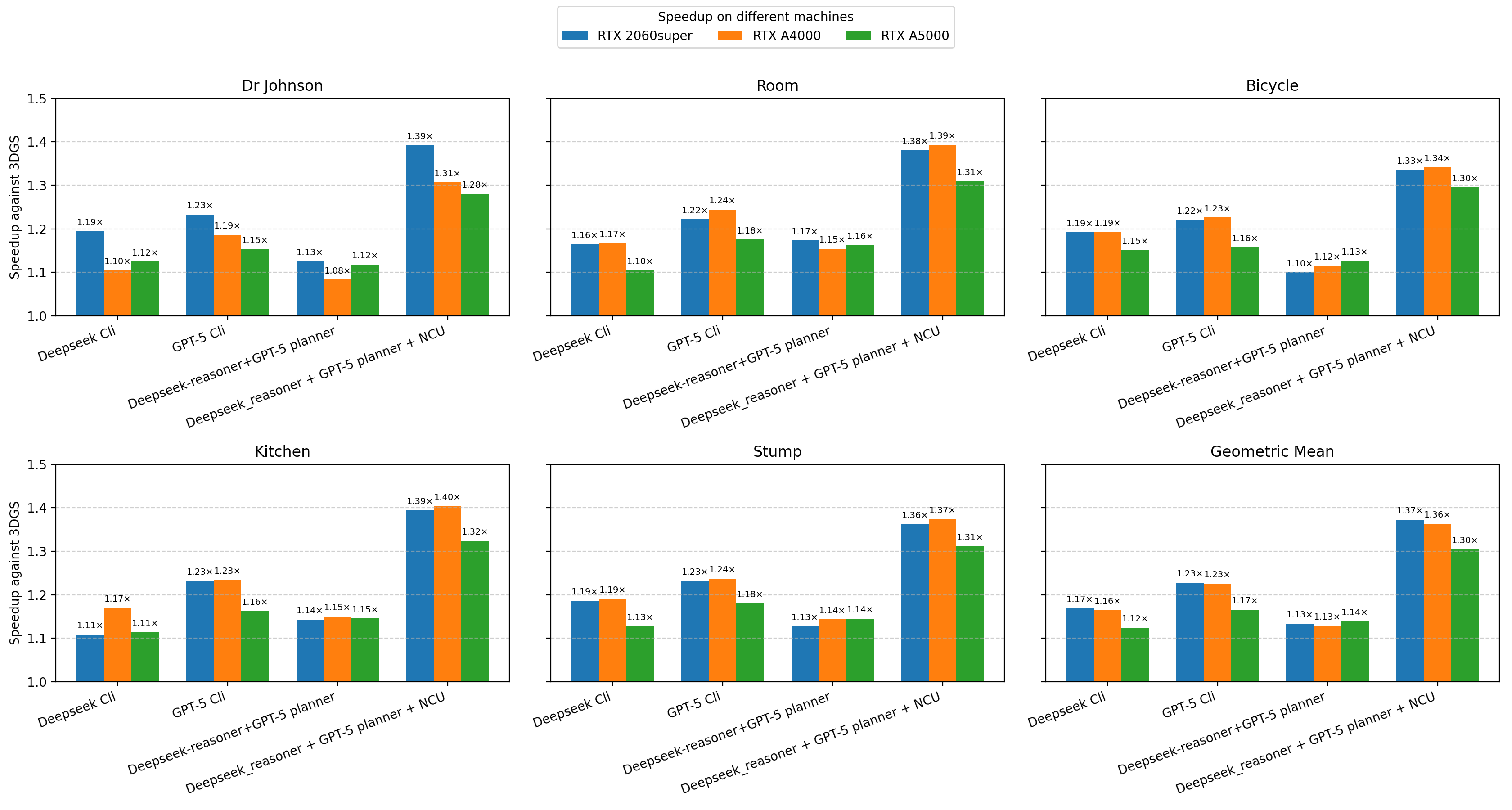}
    \caption{The speedups achieved from LLM generated optimized code against the original 3DGS kernel on different machines.}
    \label{machines}
\end{figure*}

\begin{figure}
    \centering
    \includegraphics[width=0.8\linewidth]{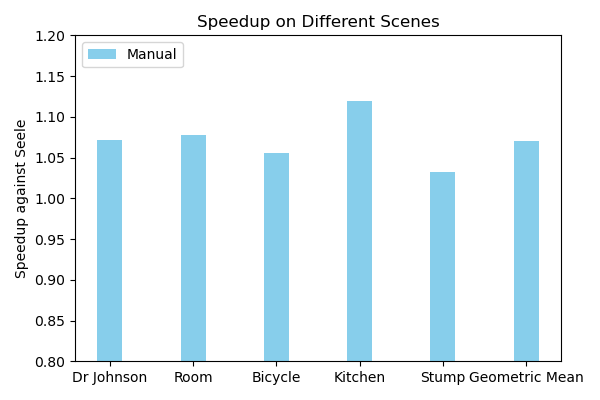}
    \caption{Evaluation of manually corrected LLM optimized code of Seele on different scenes with group size (4*8).}
    \label{general_seele}
\end{figure}


Giving that the scenario used in Evolutionary Search is limited, we further test the generality of the generated code on different scenes from Dr Johnson and MipNeRF360. As shown in Figure \ref{general_3dgs}, the average speedup from the best candidate found by Deepseek-reasoner + GPT-5 planner + Nsight Compute is 38\%, which is less than 68\% reported during the Evolutionary Search, although it is still higher than that from Deepseek-reasoner + GPT-5 planner with a speedup of 16\%. This fact implies the potential overfitting (to the input) in the generated code after a decent number of iterations of traversing the search space. The high-dimension search space and the combinations of different approaches and parameters provide the foundation for LLM to generate an overfitted version of code. In our experiments, the speedup from Deepseek-reasoner without a planner is not reliable as it failed in many views. 
Overall, the average speedup from that best candidate found by Deepseek-reasoner + GPT-5 planner + Nsight Compute is close to that of our best-effort manually optimized version with an average speedup of 39\%, which demonstrates the potential capability of LLMs. 

Figure \ref{machines} presents the speedup of LLM-generated code (using RTX 4060) relative to the baseline across three GPU machines and five scenes. Each subplot corresponds to one scene, and the final subplot reports the geometric mean across all scenes. Bars are grouped by machine with consistent coloring across subplots. The speedup from the best candidate found by Deepseek-reasoner + GPT-5 planner + Nsight Compute is at least 30\%, which highlights the portability and scalability of the proposed techniques.  

For Seele, as shown in Figure \ref{general_seele}, the average speedup of our manually corrected GPT-optimized code is 6\%. For TC-GS, no speedup is observed after we manually fix the errors in the GPT5 optimized code. 

\subsection{Limitations}
The goal of correctness check is automatic code repair of the generated code, otherwise a large number of iterations would be wasted. In this paper, we only evaluate the capability of LLMs to verify the functional equivalence of the generated code, which is the preliminary step for LLM-based code repair. Without automatic code repair, functional equivalence is not guaranteed in the generated code of Seele and TC-GS, which is the major challenge for search-based code generation. Our future work will explore ways for automatic code repair. 




\section{Conclusions}





In this paper, we present our study on analyzing and optimizing 3D Gaussian splatting pipeline using LLMs. Our evaluations demonstrate that LLMs can complement domain expertise by proposing code optimizations and exploring combined transformations through search-based code generation. However, functional equivalence of LLM generated code remains a critical challenge. 

For the original 3DGS rendering kernels, the LLM optimized code achieves up to
42\% and 38\% on average performance improvement. In comparison, our
best-effort manually optimized version can achieve a performance
improvement up to 48\% and 39\% on average. For a highly optimized and more complex 3DGS
framework, Seele, the LLM optimized code with our manual bug fixing leads to a speedup of 6\%. 



\bibliographystyle{IEEEtranS}

\begin{thebibliography}{10}
\providecommand{\url}[1]{#1}
\csname url@samestyle\endcsname
\providecommand{\newblock}{\relax}
\providecommand{\bibinfo}[2]{#2}
\providecommand{\BIBentrySTDinterwordspacing}{\spaceskip=0pt\relax}
\providecommand{\BIBentryALTinterwordstretchfactor}{4}
\providecommand{\BIBentryALTinterwordspacing}{\spaceskip=\fontdimen2\font plus
\BIBentryALTinterwordstretchfactor\fontdimen3\font minus \fontdimen4\font\relax}
\providecommand{\BIBforeignlanguage}[2]{{%
\expandafter\ifx\csname l@#1\endcsname\relax
\typeout{** WARNING: IEEEtranS.bst: No hyphenation pattern has been}%
\typeout{** loaded for the language `#1'. Using the pattern for}%
\typeout{** the default language instead.}%
\else
\language=\csname l@#1\endcsname
\fi
#2}}
\providecommand{\BIBdecl}{\relax}
\BIBdecl

\bibitem{achiam2023gpt}
J.~Achiam, S.~Adler, S.~Agarwal, L.~Ahmad, I.~Akkaya, F.~L. Aleman, D.~Almeida, J.~Altenschmidt, S.~Altman, S.~Anadkat \emph{et~al.}, ``Gpt-4 technical report,'' \emph{arXiv preprint arXiv:2303.08774}, 2023.

\bibitem{agarwal2024codemirage}
V.~Agarwal, Y.~Pei, S.~Alamir, and X.~Liu, ``Codemirage: Hallucinations in code generated by large language models,'' \emph{arXiv preprint arXiv:2408.08333}, 2024.

\bibitem{anthropic2024claude3}
\BIBentryALTinterwordspacing
{Anthropic}, ``Introducing the next generation of claude,'' \url{https://www.anthropic.com/news/claude-3-family}, Mar. 2024, accessed: 2025-09-06. [Online]. Available: \url{https://www.anthropic.com/news/claude-3-family}
\BIBentrySTDinterwordspacing

\bibitem{barron2022mip}
J.~T. Barron, B.~Mildenhall, D.~Verbin, P.~P. Srinivasan, and P.~Hedman, ``Mip-nerf 360: Unbounded anti-aliased neural radiance fields,'' in \emph{Proceedings of the IEEE/CVF conference on computer vision and pattern recognition}, 2022, pp. 5470--5479.

\bibitem{bouzenia2024repairagent}
I.~Bouzenia, P.~Devanbu, and M.~Pradel, ``Repairagent: An autonomous, llm-based agent for program repair,'' \emph{arXiv preprint arXiv:2403.17134}, 2024.

\bibitem{duan20244d}
Y.~Duan, F.~Wei, Q.~Dai, Y.~He, W.~Chen, and B.~Chen, ``4d-rotor gaussian splatting: towards efficient novel view synthesis for dynamic scenes,'' in \emph{ACM SIGGRAPH 2024 Conference Papers}, 2024, pp. 1--11.

\bibitem{gao2023pal}
L.~Gao, A.~Madaan, S.~Zhou, U.~Alon, P.~Liu, Y.~Yang, J.~Callan, and G.~Neubig, ``Pal: Program-aided language models,'' in \emph{International Conference on Machine Learning}.\hskip 1em plus 0.5em minus 0.4em\relax PMLR, 2023, pp. 10\,764--10\,799.

\bibitem{gui2024balanced}
H.~Gui, L.~Hu, R.~Chen, M.~Huang, Y.~Yin, J.~Yang, Y.~Wu, C.~Liu, Z.~Sun, X.~Zhang \emph{et~al.}, ``Balanced 3dgs: Gaussian-wise parallelism rendering with fine-grained tiling,'' \emph{arXiv preprint arXiv:2412.17378}, 2024.

\bibitem{guo2025deepseek}
D.~Guo, D.~Yang, H.~Zhang, J.~Song, R.~Zhang, R.~Xu, Q.~Zhu, S.~Ma, P.~Wang, X.~Bi \emph{et~al.}, ``Deepseek-r1: Incentivizing reasoning capability in llms via reinforcement learning,'' \emph{arXiv preprint arXiv:2501.12948}, 2025.

\bibitem{hanson2025speedy}
A.~Hanson, A.~Tu, G.~Lin, V.~Singla, M.~Zwicker, and T.~Goldstein, ``Speedy-splat: Fast 3d gaussian splatting with sparse pixels and sparse primitives,'' in \emph{Proceedings of the Computer Vision and Pattern Recognition Conference}, 2025, pp. 21\,537--21\,546.

\bibitem{hoffmann2022training}
J.~Hoffmann, S.~Borgeaud, A.~Mensch, E.~Buchatskaya, T.~Cai, E.~Rutherford, D.~de~Las~Casas, L.~Hendricks, J.~Welbl, A.~Clark \emph{et~al.}, ``Training compute-optimal large language models (2022),'' \emph{arXiv preprint arXiv:2203.15556}, 2022.

\bibitem{hollein20243dgs}
L.~H{\"o}llein, A.~Bo{\v{z}}i{\v{c}}, M.~Zollh{\"o}fer, and M.~Nie{\ss}ner, ``3dgs-lm: Faster gaussian-splatting optimization with levenberg-marquardt,'' \emph{arXiv preprint arXiv:2409.12892}, 2024.

\bibitem{hong2025autocomp}
C.~Hong, S.~Bhatia, A.~Cheung, and Y.~S. Shao, ``Autocomp: Llm-driven code optimization for tensor accelerators,'' \emph{arXiv preprint arXiv:2505.18574}, 2025.

\bibitem{hong2024llm}
C.~Hong, S.~Bhatia, A.~Haan, S.~K. Dong, D.~Nikiforov, A.~Cheung, and Y.~S. Shao, ``Llm-aided compilation for tensor accelerators,'' in \emph{2024 IEEE LLM Aided Design Workshop (LAD)}.\hskip 1em plus 0.5em minus 0.4em\relax IEEE, 2024, pp. 1--14.

\bibitem{huang2025seele}
X.~Huang, H.~Zhu, Z.~Liu, W.~Lin, X.~Liu, Z.~He, J.~Leng, M.~Guo, and Y.~Feng, ``Seele: A unified acceleration framework for real-time gaussian splatting,'' \emph{arXiv preprint arXiv:2503.05168}, 2025.

\bibitem{drjohnson_webpage}
{INRIA}, ``Dr. johnson dataset,'' \url{https://repo-sam.inria.fr/fungraph/hybrid-ibr/datasets/Dr_Johnson/index.html}, 2023, accessed: 2025-06-14.

\bibitem{kerbl20233d}
B.~Kerbl, G.~Kopanas, T.~Leimk{\"u}hler, and G.~Drettakis, ``3d gaussian splatting for real-time radiance field rendering.'' \emph{ACM Trans. Graph.}, vol.~42, no.~4, pp. 139--1, 2023.

\bibitem{lange2025ai}
R.~T. Lange, A.~Prasad, Q.~Sun, M.~Faldor, Y.~Tang, and D.~Ha, ``The ai cuda engineer: Agentic cuda kernel discovery, optimization and composition,'' Technical report, Sakana AI, 02 2025, Tech. Rep., 2025.

\bibitem{lee2024characterization}
J.~Lee, Y.~Lee, Y.~Kwon, and M.~Rhu, ``Characterization and analysis of the 3d gaussian splatting rendering pipeline,'' \emph{IEEE Computer Architecture Letters}, 2024.

\bibitem{li2025uni}
C.~Li, S.~Li, L.~Jiang, J.~Zhang, and Y.~C. Lin, ``Uni-render: A unified accelerator for real-time rendering across diverse neural renderers,'' in \emph{2025 IEEE International Symposium on High Performance Computer Architecture (HPCA)}.\hskip 1em plus 0.5em minus 0.4em\relax IEEE, 2025, pp. 246--260.

\bibitem{li2025gaurast}
S.~Li, B.~Keller, Y.~C. Lin, and B.~Khailany, ``Gaurast: Enhancing gpu triangle rasterizers to accelerate 3d gaussian splatting,'' \emph{arXiv preprint arXiv:2503.16681}, 2025.

\bibitem{liao2025tc}
Z.~Liao, J.~Ding, R.~Fu, S.~Cui, R.~Gong, L.~Wang, B.~Hu, Y.~Wang, H.~Li, X.~Zhang \emph{et~al.}, ``Tc-gs: A faster gaussian splatting module utilizing tensor cores,'' \emph{arXiv preprint arXiv:2505.24796}, 2025.

\bibitem{muller2022instant}
T.~M{\"u}ller, A.~Evans, C.~Schied, and A.~Keller, ``Instant neural graphics primitives with a multiresolution hash encoding,'' \emph{ACM transactions on graphics (TOG)}, vol.~41, no.~4, pp. 1--15, 2022.

\bibitem{novikov2025alphaevolve}
A.~Novikov, N.~V{\~u}, M.~Eisenberger, E.~Dupont, P.-S. Huang, A.~Z. Wagner, S.~Shirobokov, B.~Kozlovskii, F.~J. Ruiz, A.~Mehrabian \emph{et~al.}, ``Alphaevolve: A coding agent for scientific and algorithmic discovery,'' \emph{arXiv preprint arXiv:2506.13131}, 2025.

\bibitem{nvidia_nsight_compute_user_guide_2025}
\BIBentryALTinterwordspacing
{NVIDIA Corporation}, \emph{NVIDIA Nsight Compute User Guide}, 2025, accessed: 2025-09-12. [Online]. Available: \url{https://docs.nvidia.com/nsight-compute/NsightCompute/index.html}
\BIBentrySTDinterwordspacing

\bibitem{openai_gpt5_system_card_2025}
OpenAI, ``Gpt-5 system card,'' \url{https://openai.com/index/gpt-5-system-card/}, Aug. 2025, accessed 2025-09-11.

\bibitem{tang2025compiler}
S.~Tang, C.~Priebe, R.~Mahapatra, L.~Qin, and H.~Esmaeilzadeh, ``Compiler optimization via llm reasoning for efficient model serving,'' \emph{arXiv preprint arXiv:2506.01374}, 2025.

\bibitem{team2024gemini}
G.~Team, P.~Georgiev, V.~I. Lei, R.~Burnell, L.~Bai, A.~Gulati, G.~Tanzer, D.~Vincent, Z.~Pan, S.~Wang \emph{et~al.}, ``Gemini 1.5: Unlocking multimodal understanding across millions of tokens of context,'' \emph{arXiv preprint arXiv:2403.05530}, 2024.

\bibitem{wadhwa2024core}
N.~Wadhwa, J.~Pradhan, A.~Sonwane, S.~P. Sahu, N.~Natarajan, A.~Kanade, S.~Parthasarathy, and S.~Rajamani, ``Core: Resolving code quality issues using llms,'' \emph{Proceedings of the ACM on Software Engineering}, vol.~1, no. FSE, pp. 789--811, 2024.

\bibitem{wang2024large}
Z.~Wang, Z.~Zhou, D.~Song, Y.~Huang, S.~Chen, L.~Ma, and T.~Zhang, ``Where do large language models fail when generating code?'' \emph{arXiv e-prints}, pp. arXiv--2406, 2024.

\bibitem{zheng2020ansor}
L.~Zheng, C.~Jia, M.~Sun, Z.~Wu, C.~H. Yu, A.~Haj-Ali, Y.~Wang, J.~Yang, D.~Zhuo, K.~Sen \emph{et~al.}, ``Ansor: Generating $\{$High-Performance$\}$ tensor programs for deep learning,'' in \emph{14th USENIX symposium on operating systems design and implementation (OSDI 20)}, 2020, pp. 863--879.

\end{thebibliography}

\clearpage
\appendices

\section{Examples of errors in LLM-generated code of TC-GS}

\begin{figure}[H]
    \centering
    \includegraphics[width=0.9\linewidth]{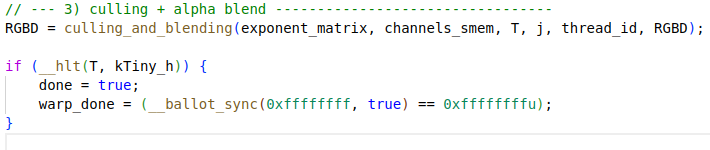}
    \caption{Case of an error in the generated code.}
    \label{tcgserror}
\end{figure}
\begin{figure}[H]
    \centering
    \includegraphics[width=0.9\linewidth]{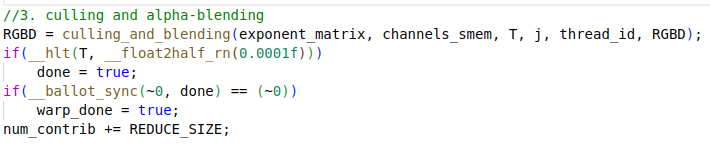}
    \caption{The original version in TC-GS}
    \label{tcgscorrect}
\end{figure}

Here we demonstrate the error in the LLM-generated code of TC-GS. As shown in Figure \ref{tcgserror}, \textit{\_\_ballot\_sync} is called inside a if-statement with a full mask 0xffffffff. However, due to the divergence brought in by that if-statement, the full active mask at this stage is not guaranteed. On the contrary, in the original code shown in Figure \ref{tcgscorrect}, \textit{\_\_ballot\_sync} is called outside that if-statement after convergence.   

\section{Examples of code generation for 3DGS}

Here we provide some examples of our code generation. The following code is the best candidate generated by Deepseek after 40 iterations without suggestions from planner.

\begin{WideVerb}
auto block = cg::this_thread_block();
const uint32_t horizontal_blocks = (W + BLOCK_X - 1) / BLOCK_X;
const uint2 pix = {
    block.group_index().x * BLOCK_X + block.thread_index().x,
    block.group_index().y * BLOCK_Y + block.thread_index().y
};
if (pix.x >= W || pix.y >= H) return;

const uint32_t pix_id = W * pix.y + pix.x;
const float2 pixf = {(float)pix.x, (float)pix.y};
const uint2 range = ranges[block.group_index().y * horizontal_blocks + block.group_index().x];
const int num_points = range.y - range.x;
const int rounds = (num_points + BLOCK_SIZE - 1) / BLOCK_SIZE;

__shared__ int point_ids[BLOCK_SIZE];
__shared__ float2 points_xy[BLOCK_SIZE];
__shared__ float4 conic_opacities[BLOCK_SIZE];

float T = 1.0f;
uint32_t last_contributor = 0;
float color_accum[CHANNELS] = {0};
float invdepth_accum = 0.0f;
bool compute_depth = (invdepth != nullptr);

for (int round = 0; round < rounds; round++) {
    int load_idx = round * BLOCK_SIZE + block.thread_rank();
    int point_idx = range.x + load_idx;
    
    if (point_idx < range.y) {
        uint32_t pid = point_list[point_idx];
        point_ids[block.thread_rank()] = pid;
        points_xy[block.thread_rank()] = points_xy_image[pid];
        conic_opacities[block.thread_rank()] = conic_opacity[pid];
    }
    block.sync();

    int points_to_process = min(BLOCK_SIZE, num_points - round * BLOCK_SIZE);
    for (int i = 0; i < points_to_process; i++) {
        float2 xy = points_xy[i];
        float2 delta = {xy.x - pixf.x, xy.y - pixf.y};
        float4 con_o = conic_opacities[i];
        
        // Use named variables for better readability
        float a = con_o.x, b = con_o.y, c = con_o.z, opacity = con_o.w;
        float dx2 = delta.x * delta.x;
        float dy2 = delta.y * delta.y;
        float dxy = delta.x * delta.y;
        float power = -0.5f * (a * dx2 + c * dy2) - b * dxy;
        
        if (power > 0.0f) continue;

        float alpha = min(0.99f, opacity * __expf(power));
        if (alpha < 1.0f / 255.0f) continue;
        
        float new_T = T * (1.0f - alpha);
        if (new_T < 0.0001f) break;

        int pid = point_ids[i];
        const float* feature_ptr = features + pid * CHANNELS;
        float weight = alpha * T;
        
        // Unroll the channel loop for better performance
        #pragma unroll
        for (int ch = 0; ch < CHANNELS; ch++) {
            color_accum[ch] += feature_ptr[ch] * weight;
        }

        if (compute_depth) {
            invdepth_accum += __frcp_rn(depths[pid]) * weight;
        }

        T = new_T;
        last_contributor = load_idx + i + 1;
    }
    block.sync();
}

final_T[pix_id] = T;
n_contrib[pix_id] = last_contributor;

// Unroll the channel loop for better performance
#pragma unroll
for (int ch = 0; ch < CHANNELS; ch++) {
    out_color[ch * H * W + pix_id] = color_accum[ch] + T * bg_color[ch];
}

if (compute_depth) {
    invdepth[pix_id] = invdepth_accum;
}
\end{WideVerb}

An Example of the prompt for Openevolve framework.

\begin{WideVerb}

"You are an expert coder helping to improve programs through evolution.
Rewrite the code inside the EVOLVE-BLOCK. 
Return the full code, but keep EVOLVE-BLOCK markers. 
Do not change anything else in the file. 
Here are first 5 to try. 
Shared broadcast of features/depths per j. 
Warp-level early exit & warp-scope sync. 
FMA + fast-math. 
Template & unroll CHANNELS. 
Order contributors to hit T cutoff earlier."

\end{WideVerb}

The following code is the best candidate generated by Deepseek after 40 iterations with the above prompt.

\begin{WideVerb}
auto block = cg::this_thread_block();
auto warp = cg::tiled_partition<WARP_SIZE>(block);
uint32_t horizontal_blocks = (W + BLOCK_X - 1) / BLOCK_X;
uint2 pix_min = { block.group_index().x * BLOCK_X, block.group_index().y * BLOCK_Y };
uint2 pix_max = { min(pix_min.x + BLOCK_X, W), min(pix_min.y + BLOCK_Y , H) };
uint2 pix = { pix_min.x + block.thread_index().x, pix_min.y + block.thread_index().y };
uint32_t pix_id = W * pix.y + pix.x;
float2 pixf = { (float)pix.x, (float)pix.y };

bool inside = pix.x < W&& pix.y < H;
bool done = !inside;

uint2 range = ranges[block.group_index().y * horizontal_blocks + block.group_index().x];
const int rounds = ((range.y - range.x + BLOCK_SIZE - 1) / BLOCK_SIZE);
int toDo = range.y - range.x;

__shared__ int collected_id[BLOCK_SIZE];
__shared__ float2 collected_xy[BLOCK_SIZE];
__shared__ float4 collected_conic_opacity[BLOCK_SIZE];
__shared__ float collected_features[BLOCK_SIZE][CHANNELS];
__shared__ float collected_depths[BLOCK_SIZE];

float T = 1.0f;
uint32_t contributor = 0;
uint32_t last_contributor = 0;
float C[CHANNELS] = { 0 };
float expected_invdepth = 0.0f;

for (int i = 0; i < rounds; i++, toDo -= BLOCK_SIZE)
{
    if (warp.all(done)) break;

    int progress = i * BLOCK_SIZE + block.thread_rank();
    if (range.x + progress < range.y)
    {
        int coll_id = point_list[range.x + progress];
        collected_id[block.thread_rank()] = coll_id;
        collected_xy[block.thread_rank()] = points_xy_image[coll_id];
        collected_conic_opacity[block.thread_rank()] = conic_opacity[coll_id];
        collected_depths[block.thread_rank()] = depths ? __frcp_rn(depths[coll_id]) : 0.0f;
        #pragma unroll
        for (int ch = 0; ch < CHANNELS; ch++) {
            collected_features[block.thread_rank()][ch] = features[coll_id * CHANNELS + ch];
        }
    }
    block.sync();

    #pragma unroll 4
    for (int j = 0; j < min(BLOCK_SIZE, toDo); j++)
    {
        if (done) continue;
        contributor++;

        float2 xy = collected_xy[j];
        float2 d = { xy.x - pixf.x, xy.y - pixf.y };
        float4 con_o = collected_conic_opacity[j];
        
        float dx2 = d.x * d.x;
        float dy2 = d.y * d.y;
        float dxy = d.x * d.y;
        float power = fmaf(-0.5f, fmaf(con_o.x, dx2, con_o.z * dy2), -con_o.y * dxy);
        
        if (power > 0.0f) continue;
        float alpha = min(0.99f, con_o.w * __expf(power));
        if (alpha < 1.0f / 255.0f) continue;
        
        float test_T = __fmul_rn(T, 1.0f - alpha);
        if (test_T < 0.0001f) {
            done = true;
            continue;
        }

        float alpha_T = __fmul_rn(alpha, T);
        
        #pragma unroll
        for (int ch = 0; ch < CHANNELS; ch++) {
            C[ch] = __fmaf_rn(collected_features[j][ch], alpha_T, C[ch]);
        }

        if (invdepth) {
            expected_invdepth = __fmaf_rn(collected_depths[j], alpha_T, expected_invdepth);
        }

        T = test_T;
        last_contributor = contributor;
    }

    if (warp.all(done)) break;
}

if (inside)
{
    final_T[pix_id] = T;
    n_contrib[pix_id] = last_contributor;
    
    #pragma unroll
    for (int ch = 0; ch < CHANNELS; ch++) {
        out_color[ch * H * W + pix_id] = __fmaf_rn(T, bg_color[ch], C[ch]);
    }

    if (invdepth) {
        invdepth[pix_id] = expected_invdepth;
    }
}
\end{WideVerb}

\end{document}